\documentclass[a4paper]{article}
\usepackage{Odyssey2022}
\usepackage{epsfig,amssymb,amsmath}
\ninept
\usepackage{hyperref}
\usepackage{lineno,multirow,cite}
\usepackage{multirow}
\usepackage{multicol}
\usepackage{amsmath} 
\usepackage{enumitem}
\usepackage{color}
\usepackage{comment}
\usepackage{booktabs}

\usepackage[caption=false,font=footnotesize]{subfig}
\usepackage{tabularx}

\title{Language-Independent Speaker Anonymization Approach \\ using Self-Supervised Pre-Trained Models}

\name{Xiaoxiao Miao$^1$, Xin Wang$^1$, Erica Cooper$^1$, Junichi Yamagishi$^{1}$, Natalia Tomashenko$^2$}
\address{
$^1$National Institute of Informatics, Japan 
 $^2$LIA, University of Avignon, France \\
{\small \tt xiaoxiaomiao@nii.ac.jp}}

\begin{document}
\maketitle
\begin{abstract}
Speaker anonymization aims to protect the privacy of speakers while preserving spoken linguistic information from speech. Current mainstream neural network speaker anonymization systems are complicated, containing an F0 extractor, speaker encoder, automatic speech recognition acoustic model (ASR AM), speech synthesis acoustic model and speech waveform generation model. 
Moreover, as an ASR AM is language-dependent, trained on English data, it is hard to adapt it into another language. 
In this paper, we propose a simpler self-supervised learning (SSL)-based method for language-independent speaker anonymization without any explicit language-dependent model, which can be easily used for other languages. Extensive experiments were conducted on the VoicePrivacy Challenge 2020 datasets in English and AISHELL-3 datasets in Mandarin to demonstrate the effectiveness of our proposed SSL-based language-independent speaker anonymization method\footnote{English and Mandarin Audio samples and source code are available at \url{https://github.com/nii-yamagishilab/SSL-SAS}}.
\end{abstract}

\section{Introduction}
\label{sec:intro}
It is well known that speech data contains a plethora of privacy information, such as speaker identity, age, emotion, gender, and so on.
The speaker identities of speech recordings without any protection could be resynthesized, cloned, and converted by using advanced speech synthesis (SS) technologies, which may lead to a privacy risk \cite{vestman2020voice}. 
In fact, speaker anonymization is a way to conceal speaker information while maintaining intelligibility and naturalness as much as possible \cite{jin2008voice, fang2019speaker, tomashenko2020introducing,tomashenko2021voiceprivacy}.

In recent decades, voice protections have mainly focused on noise addition, voice transformation, conversion, synthesis and disentangled representation learning \cite{7472729,qian2017voicemask, jin2009voice, qian2018hidebehind, huang2021defending, magarinos2017reversible, fang2019speaker,srivastava2019privacy}.
Efforts to study the trade-off between speaker identity and speech intelligibility by adding noise to test data have also been made.
The goal of voice transformation, conversion, synthesis, and disentangled representation learning methods is to synthesize various speakers' voices. 
This can be achieved by modifying the non-linguistic information in speech, such as F0, energy, or speaker vector, or by producing a new speaker voice with generative models.

To define the speaker anonymization problem accurately and fairly, the VoicePrivacy Challenge (VPC) 2020 \cite{tomashenko2020introducing,tomashenko2021voiceprivacy} has provided common datasets, evaluation metrics (objective and subjective), and two main baselines to suppress the individuality of speech, leaving other speaker attribute information such as age and gender, while preserving the naturalness of speech.
The primary baseline of VPC 2020 consists of several components: an F0 extractor, pre-trained time delay neural network (TDNN) x-vector automatic speaker verification (ASV) system \cite{snyder2018x}, pre-trained factorized time delay neural network (TDNN-F) automatic speech recognition acoustic model (ASR AM) \cite{povey2018semi, peddinti2015time}, speech synthesis acoustic model (SS AM), and neural source-filter (NSF) \cite{wang2019neural} waveform model, denoted as B1. 
Although the system can effectively anonymize the speech data, it is complicated to build and deploy.
Moreover, the ASR AM of this system is language-dependent and requires language-specific resources. The system cannot be directly applied to another language.
The second baseline, which is denoted as B2, is based on conventional signal processing techniques and does not require training data. However, it has been demonstrated that B2 cannot protect the speaker information as well as the neural-network-based systems \cite{tomashenko2020introducing,tomashenko2021voiceprivacy}.

Recent success in self-supervised learning (SSL) has shown remarkable performance for speech synthesis tasks \cite{lakhotia2021generative, polyak2021speech,huang2021any}, 
One of the new speech resynthesis works \cite{polyak2021speech} learned discrete speech units from the unsupervised clustering of speech representations using a large unlabeled speech corpus. The idea of discretized speech units is to disentangle speech representations, separating content from speaker information.
In addition, a number of researchers \cite{choi2021neural,van2021comparison} have found that incorrect quantization predictions from discretization will cause mispronunciation problems.

In this paper, we propose a new SSL-based language-independent neural speaker anonymization method.
We first introduce soft content representations \cite{van2021comparison} instead of discretized ones to mitigate ambiguous mispronunciations in the anonymized speech.
Next, as VPC 2020 has shown that x-vectors are important to encode speaker identity information for the speaker anonymization task,
we update the most frequently used d-vector \cite{polyak2021speech, heigold2016end} or TDNN x-vector speaker encoders \cite{snyder2018x, tomashenko2020introducing,tomashenko2021voiceprivacy} to the state-of-the-art ECAPA-TDNN \cite{desplanques2020ecapa}.
More importantly, the current VPC 2020 primary baseline B1 requires large amounts of language-specific resources to train a language-dependent ASR AM,
while the SSL-based soft content encoder of the proposed method learns universal representations by training with unlabeled audio data, which improves the portability to a new language.  
Extensive experiments were conducted on the VPC 2020 datasets in English and AISHELL-3 \cite{shi21c_interspeech} datasets in Mandarin to demonstrate the effectiveness of our proposed SSL-based language-independent speaker anonymization method. 

In the remainder of this paper, Section \ref{sec:related work} overviews the speaker anonymization baseline systems of VPC 2020.
Section \ref{sec:ssl-based method} describes our proposed SSL-based language-independent speaker anonymization method, Section \ref{sec:eval} evaluates the system, and  
Section \ref{sec:conclusion} concludes the paper.

\begin{figure}[t]
\centering
\includegraphics[width=1.0\columnwidth]{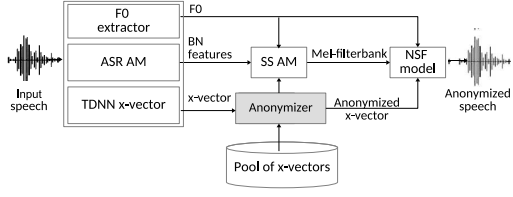}
\vspace{-5mm}
\caption{Architecture of the VPC 2020 baseline B1 system.}
\label{fig:structures}
\vspace{-3mm}
\end{figure}

\section{Speaker Anonymization Baselines}
\label{sec:related work}
\subsection{Baseline using x-vector and Neural Waveform Models}
The VPC 2020 primary baseline aims to decompose speech into speaker identity, fundamental frequency, and linguistic information, which is illustrated in Figure \ref{fig:structures} and is composed of three main procedures:

\noindent
\textit{1) Original F0, bottleneck (BN), and x-vector extraction.}
There are two pre-trained and fixed encoders, 
including a TDNN-F ASR AM \cite{povey2018semi, peddinti2015time} trained on the \textit{LibriSpeech-train-clean-100} and \textit{LibriSpeech-train-other-500} datasets  \cite{panayotov2015librispeech} to extract 256-dimensional BN features as content representation, and a
TDNN x-vector model trained on the \textit{VoxCeleb-1 \& 2} \cite{nagrani2017voxceleb, chung2018voxceleb2} datasets to extract 512-dimensional x-vector as the speaker identity representation.

\noindent
\textit{2) x-vector anonymization.}
A pseudo x-vector is obtained by searching the 200 furthest same-gender x-vectors from an external x-vector pool (\textit{LibriTTS-train-other-500} \cite{zen2019libritts}), according to probabilistic  linear  discriminant analysis (PLDA) \cite{snyder2018x} distances. Then averaging 100 randomly-selected ones\cite{Srivastava2020DesignCF}.

\noindent
\textit{3) Anonymized speech synthesis.}
An SS AM first generates a Mel-filterbank from the anonymized x-vector, original F0, and BN features. The generated Mel-filterbank, F0, and anonymized x-vector are fed into an NSF \cite{wang2019neural} model to synthesize the anonymized waveform.
Both SS AM and NSF models are trained with \textit{LibriTTS-train-clean-100} \cite{zen2019libritts}.

\subsection{ Baseline using McAdams Coefficient}
In short-time speech analysis, the power spectrum of a short-term segment of speech is fit with an all-pole model using linear predictive coding analysis.
The formant frequencies are determined by the angles of the corresponding complex poles.
It is known that the McAdams coefficient can be used to alter the pole angles  \cite{mcadams1984spectral,patino2020speaker}. Thus, in the second baseline of VPC 2020, the speaker anonymization process was achieved by varying the pole angles to shift the formants using the McAdams coefficient.

\section{Proposed Method using SSL Models}
\label{sec:ssl-based method}

\begin{figure}[t]
\centering
\includegraphics[width=1.0\columnwidth]{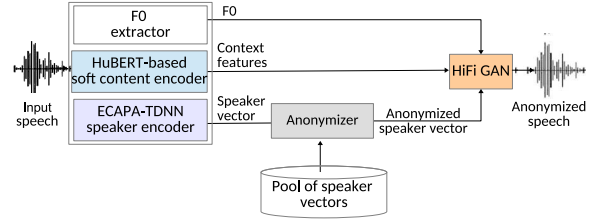}
\vspace{-5mm}
\caption{Architecture of the proposed system. The main differences between the B1 system and proposed system are highlighted in color.}
\label{fig:structure-proposed}
\vspace{-5mm}
\end{figure}

\begin{figure}[t]
\vspace{5mm}
\centering\includegraphics[width=0.95\columnwidth]{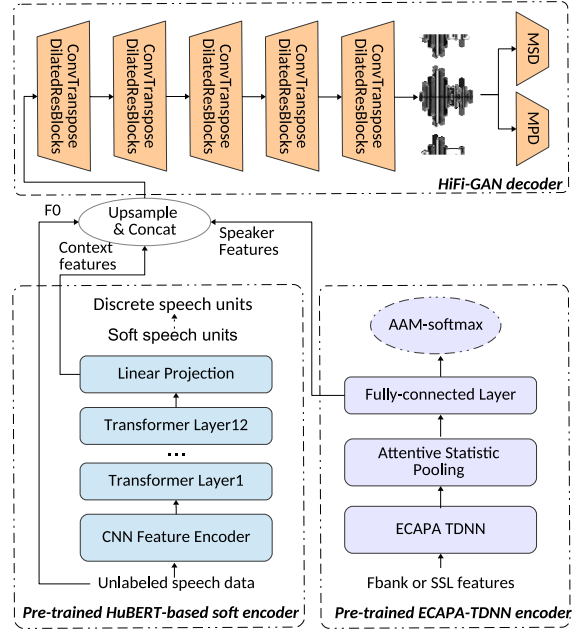}
\vspace{-5mm}
\caption{Language-independent speaker anonymization system using SSL models showing expanded detail of the combined structure.}
\label{fig:proposed-system-details}
\vspace{-5mm}
\end{figure}

In this section, we describe the proposed SSL-based language-independent speaker anonymization system as illustrated in Figure~\ref{fig:structure-proposed} and a more detailed structure is shown in  Figure~\ref{fig:proposed-system-details}.
We can see the differences clearly between B1 and the proposed SSL-based system from Figure \ref{fig:structure-proposed}, which consists of two pre-trained and fixed encoders: a HuBERT-based soft content encoder $ E_c $, ECAPA-TDNN speaker encoder $ E_{spk} $, one F0 extractor and one decoder HiFi-GAN neural vocoder \cite{kong2020hifi}. These approaches have been successfully applied to the different tasks. We are the first to reassemble them for the purpose of the speaker anonymization task.

The HuBERT-based soft content encoder \cite{van2021comparison} is obtained by fine-tuning a pretrained HuBERT Base model \cite{hsu2021hubert},
learning finer grained continuous context representations from discrete units. 
The ECAPA-TDNN speaker encoder provides an utterance-level discriminative speaker-related representation.
We follow the same x-vector anonymization scheme of the VPC baseline B1 system \cite{tomashenko2020introducing,tomashenko2021voiceprivacy} to anonymize a speaker vector of each source input utterance.
For the F0 extractor, the YAAPT algorithm \cite{kasi2002yet} is used to extract F0 from the input signal.
Then the content features, F0 and anonymized speaker vector are fed into the HiFi-GAN neural vocoder after upsampling and concatenating to generate the anonymized speech.

\subsection{HuBERT-based Soft Content Encoder}
\label{sec:hubert soft}
Many self-supervised speech models have been proposed in previous studies. 
It has been observed that different layers of SSL-based models contain different information like speaker identity, content and semantics \cite{pasad2021layer,baevski2020wav2vec,chen2021large,yang21c_interspeech}. 
In particular, higher and middle layers of SSL-based models tend to capture richer linguistic information.
Replacing the typical ASR bottleneck features with SSL-based content features has the advantage of eliminating the need for ASR resources such as phone labels, phonesets, lexicons, and language models.

This study mainly focuses on HuBERT \cite{hsu2021hubert}.
It was trained on the partially masked frames using pseudo labels, which can be obtained and refined by iterative clustering processes.
There are two common ways to use a pre-trained HuBERT model in speech synthesis. 
One is directly using the continuous  output features of the HuBERT model \cite{yang21c_interspeech}.
However, it has been proven that the continuous representations contain both context and speaker information, and are thus not suitable for speech disentanglement.
Another work \cite{polyak2021speech} applied a $k$-means algorithm over HuBERT continuous representations to obtain discrete speech units, getting rid of speaker identity attributes. While inaccurate discrete features are known to lead to incorrect pronunciation, and we confirm the issue in our experiments later as well.

\begin{figure}[t]
\centering\includegraphics[width=0.8\columnwidth]{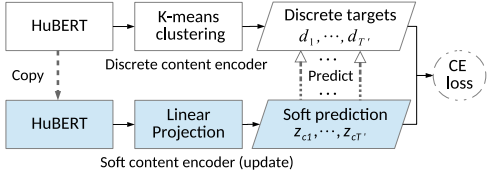}
\vspace{-3mm}
\caption{Outline of HuBERT-based soft content encoder.}
\label{fig:soft content}
\vspace{-5mm}
\end{figure}

To compromise a trade-off between continuous  representations and discrete speech units, we implement the HuBERT-based soft content encoder that is similar to \cite{van2021comparison} to capture soft content features by predicting a distribution over discrete speech units.
These soft content features are expected to represent more accurate content information while suppressing speaker information effectively. 
Figure \ref{fig:soft content} presents the training process of the HuBERT-based soft content encoder.
Given a sequence of an input recording  $ \mathbf{x} = (x_1, \cdots, x_T)$, where $ x_t \in \mathcal{R} $, the output of the soft content $ E_c $ is the sequence of soft prediction $ \mathbf{z}_c = ( \mathbf{z}_{c1}, \cdots, \mathbf{z}_{cT'})$, 
where $ T' / T = 1/320 $ is determined by the convolutional neural network (CNN) stride of HuBERT. 
During the training stage, given an input utterance,
the pre-trained HuBERT with a $k$-means clustering model is firstly introduced as the fixed model to extract discrete speech units $( d_{1}, \cdots, d_{T'})$ as targets. 
Then, the $k$-means clustering strategy of the pre-trained HuBERT is replaced with a linear projection layer that transforms the HuBERT output to a sequence of soft prediction $ \mathbf{z}_c = ( \mathbf{z}_{c1}, \cdots, \mathbf{z}_{cT'})$.
The parameterized distribution of each soft unit over the dictionary of discrete speech units can be defined in Equation \ref{eqn:soft encoder}:
\begin{eqnarray} \label{eqn:soft encoder}
p(d_{t'} = i \mid \mathbf{z}_{ct'}) = \frac{\exp(\text{sim}(\mathbf{z}_{ct'}, \mathbf{w}_i) / \tau)}{\sum_{k=1}^K \exp(\text{sim}(\mathbf{z}_{ct'}, \mathbf{w}_k) / \tau)},
\end{eqnarray}
where $i$ denotes the cluster index of the $i^\textrm{th}$ discrete target, $\mathbf{w}_i$ is a corresponding trainable embedding vector, 
$\text{sim}(\cdot, \cdot)$ measures the cosine similarity, and $\tau$ is a temperature parameter set as 0.1 by the experimental selection.
The final step is to update the soft content (including HuBERT) by minimizing the cross-entropy (CE) loss function, which measures the distance between the distribution and discrete targets $( d_{1}, \cdots, d_{T'})$.

\subsection{ECAPA-TDNN Speaker Encoder}
\label{sec-ecapa-spk-encoder}
Current neural network ASV systems mainly consist of input feature extraction, a deep frame-level feature extractor, pooling layer, and loss function.
For input feature extraction, the most commonly used features are Mel-filterbank (FBank) and Mel-frequency cepstrum coefficients (MFCC).
Recent works have shown the representations derived from different layers of SSL-based models also contain different levels of discriminative speaker-related information \cite{chen2021large}. Specifically, lower layers of SSL-based models have more speaker information than higher layers.
For the deep feature extractor, TDNN-based \cite{snyder2018x,desplanques2020ecapa} and CNN-based \cite{cai2018exploring,miao2021d} systems have been successfully used in the field of ASV.

In this paper, we employ the state-of-the-art ECAPA-TDNN \cite{desplanques2020ecapa} as speaker encoder $E_{s}$ to encode speaker identity information, $\mathbf{z}_{spk} = E_{s}(\mathbf{x})$, $\mathbf{z}_{spk} \in {\mathbb{R}^{ 192}}$.
ECAPA-TDNN is an improved version of the original TDNN x-vector. The deep frame-level feature extractor of ECAPA-TDNN includes squeeze-excitation blocks, skip connections, and multi-scale/multi-layer feature aggregation to explore discriminative speaker features and a channel-dependent attentive statistics pooling layer to capture fixed-dimensional representation from variable-length features.
In addition to using the commonly used FBank as input features, we also explore the effect of SSL-based input features for the ECAPA-TDNN in the proposed speaker anonymization framework.

\subsection{ F0 Extractor }
We use the YAAPT algorithm \cite{kasi2002yet} to extract F0 from the input signal $\mathbf{x}$.
Its F0 sequence is denoted as $\mathbf{z}_f = ( \mathbf{z}_{f1}, \cdots, \mathbf{z}_{fT''})$. The ratio of $ T'' / T = 1/160 $ in our experiment.

\subsection{ HiFi-GAN Neural Vocoder }
The frame-wise context $\mathbf{z}_c$ and F0 $\mathbf{z}_f$ sequences are up-sampled and concatenated. 
The segmental-level speaker embedding $\mathbf{z}_{spk}$ is then integrated to each frame in the up-sampled sequence,
namely the encoded representation $\mathbf{z} = (\mathbf{z}_c, \mathbf{z}_f, \mathbf{z}_{spk})$,
which is then passed to the neural vocoder HiFi-GAN \cite{kong2020hifi} to generate the speech waveforms.

Unlike the two-stage pipeline strategy in VPC 2020 baseline B1, there is no SS AM before the neural vocoder. 
The reason for excluding SS AM without generating a Mel-filterbank further is that the content features derived from the HuBERT-based soft context encoder contain rich linguistic information, compared with BN features extracted by ASR AM.
In addition, HiFi-GAN is known as a non autoregressive model that has been successfully applied to speech synthesis, achieving both high computational efficiency and sample quality.

HiFi-GAN consists of a generator and two discriminators. The generator has five groups of ResBlock where multiple transposed convolutions upsample low-frequency encoded representation $\mathbf{z}$ to the original audio size, followed by a stack of dilated residual connections to increase the receptive field.
During training, the generated audio sample $\hat{\mathbf{x}}$, is passed to two discriminators. 
The multi-period discriminator (MPD) captures the periodic patterns of audio through five-period sub-discriminators operating on equally spaced samples between the original and generated waveforms.
The multi-scale discriminator (MSD) has the advantage of exploring long-range and consecutive interactions of audio by multi-scale average pooling operations. 
The configuration of HiFi-GAN is the same as \cite{polyak2021speech}.
The overall training losses of HiFi-GAN involve a generator loss $\mathcal{L}_{G}$ and a discriminator loss $\mathcal{L}_{D}$:
\begin{align}
\begin{split}
 \mathcal{L}_{G} &=  \sum_{k=1}^{K}\Bigg[\mathcal{L}_{Adv}(G; D_k) + \lambda_{fm}\mathcal{L}_{FM}(G; D_k)\Bigg] \\
                    & + \lambda_{mel}\mathcal{L}_{Mel}(G) 
\end{split}
\end{align}
\begin{align}
    \mathcal{L}_{D} &= \sum_{k=1}^{K}\mathcal{L}_{Adv}(D_k; G)
\end{align}
where $D_k$ denotes the $k$-th sub-discriminator in the MPD and MSD. 
In the experiment, we set $\lambda_{fm}=2$ and $\lambda_{mel}=45$ to balance the adversarial losses,  the feature matching loss  $\mathcal{L}_{FM}$, and mel-spectrogram loss $\mathcal{L}_{Mel}$:
\begin{align}
    \mathcal{L}_{FM}(G; D_k) &= \mathbb{E}_{(\mathbf{x}, \hat{\mathbf{x}})} \Bigg[\sum_{i=1}^{L}\frac{|D_k^i(\mathbf{x})-D_k^i(G(\hat{\mathbf{x}}))|_{1}}{N_{i}}\Bigg] \\
    \mathcal{L}_{Mel}(G) &= \mathbb{E}_{(\mathbf{x}, \hat{\mathbf{x}})} \Bigg[|\phi(\mathbf{x})-\phi(G(\hat{\mathbf{x}}))|_{1}\Bigg]
\end{align}
where $L$ denotes the number of layers in the discriminator. $D^i_k$ is the $i$-th layer feature map of the $k$-th sub-discriminator, and $N_i$ indicates the number of units in the $i$-th layer.
 $\phi$ is a spectral operator computing a mel-spectrogram from a given waveform.

\section{Evaluation}
\label{sec:eval}
In this section, we first test our ECAPA-TDNN-based speaker encoders separately from the anonymization task, using the well explored Voxceleb dataset \cite{chung2018voxceleb2,nagrani2017voxceleb} to show that they have reasonable performance.
Then we followed the VPC 2020 evaluation plan \cite{tomashenko2020introducing,tomashenko2021voiceprivacy} to demonstrate that the proposed anonymization system can achieve comparable performance to VPC 2020 baselines on English data.
Finally, we conducted anonymization experiments on Mandarin data to show that the proposed system without any explicit language-dependent component can be applied to a different language.

\subsection{ASV Experiments}
\label{sec:asv}
As described in Section~\ref{sec-ecapa-spk-encoder}, we built two ECAPA-TDNN-based speaker encoders that can be plugged into our anonymization system. Both of them followed the recipe in the original ECAPA literature \cite{desplanques2020ecapa}, but their input features are different: 
\begin{itemize}
\item F-ECAPA: 80-dimensional FBank with 25ms window size and 10ms frame shift.
\item S-ECAPA: 768-dimensional weighted average of the output features from all the hidden layers of a pre-trained HuBERT Base model released by Fairseq  toolkit{\footnote{\label{fairseq}{\url{https://github.com/pytorch/fairseq/}}}}. The parameters of HuBERT Base were updated when training S-ECAPA. 
\end{itemize}
F-ECAPA used the same network typology as the original ECAPA with 512 channels in the convolution frame layers \cite{desplanques2020ecapa}. The training loss was the additive angular margin (AAM) loss \cite{deng2019arcface, xiang2019margin} with a margin of 0.2. S-ECAPA also used the same typology except for the first layer due to the different input feature dimensions. 
Both models used the 192-dimensional output from the last linear projection layer as the speaker embedding.

We trained both models using the Voxceleb2 dev set \cite{chung2018voxceleb2}, which contains over 1 million utterances from 5,994 speakers. 
Each audio sample was cropped to segments with a maximum duration of 3s during training.
Extra room impulse response data\footnote{\url{https://www.openslr.org/28/}} was used for training data augmentation, while voice activity detection (VAD) was not used. 
We then solely evaluated the trained models on the Voxceleb1 test set \cite{nagrani2017voxceleb} for the ASV task, which contains 4,872 utterances from 40 speakers. The evaluation metrics include the equal error rate (EER) and the minimum of the decision cost function (minDCF) calculated using $C_{FA} = C_{Miss} = 1$, and $ P_{target} = 0.01$, as in \cite{desplanques2020ecapa}.
This ASV experiment was conducted using the SpeechBrain toolkit \cite{speechbrain}.

\begin{table}[t]
\caption{ECAPA-TDNN-based speaker encoder included in the proposed system was solely evaluated for the ASV task. The speaker encoders were trained on the Voxceleb2 dev set and evaluated on the Voxceleb1 test set. Lower EER is better for the ASV task.}
\begin{center}
\footnotesize
\begin{tabular}{l|c|c}
\toprule
 \textbf{system}&  \textbf{EER} [\%]  & \textbf{minDCF} \\
 \hline
  TDNN x-vector~\cite{xiang2019margin}   & 2.23 & - \\  
 ECAPA \cite{desplanques2020ecapa} & 1.01  & 0.127 \\ 
\hline
 F-ECAPA & 1.10  & 0.135 \\ 
S-ECAPA & 0.87  & 0.123\\
\bottomrule
 \end{tabular}
\end{center}
\label{tab:asv}
\vspace{-5mm}
\end{table}

Table~\ref{tab:asv} compares the EER and minDCF results from our speaker encoders with those reported in the literature on TDNN x-vector and ECAPA.
Our F-ECAPA performed similarly to the original ECAPA and outperformed the TDNN x-vector by a large margin.
Compared with F-ECAPA, the S-ECAPA using the input features extracted from the HuBERT model achieved better results.
Since both F-ECAPA and S-ECAPA outperformed the TDNN xvector, which was used in the VPC 2020 primary baseline, we decided to use the two new speaker encoders in the following speaker anonymization experiments.

\begin{figure}[t]
\centering\includegraphics[width=0.9\columnwidth]{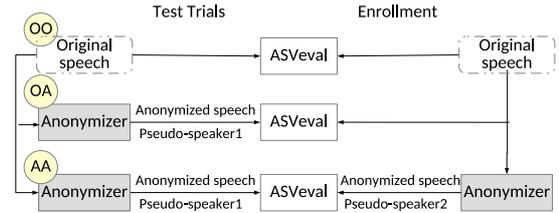}
\vspace{-3mm}
\caption{Objective speaker verifiability evaluation scenarios.}
\label{fig:eval_types}
\vspace{-5mm}
\end{figure}

\begin{table*}[thbp]
\vspace{-5mm}
\caption{ EER (\%) on English data (development set of VPC 2020) evaluated by \emph{$ASV_\text{eval}^{\textrm{\textrm{eng}}}$}.
The EER results of the OO condition for Libri-female, Libri-male, VCTK-diff-female, and VCTK-diff-male are 8.66\%, 1.24\%, 2.86\%, and 1.43\%, respectively. Higher EER indicates better privacy.}
\vspace{1mm}
\label{tab:vpc-dev-eer}
\setlength{\tabcolsep}{12pt}
  \centering
  \footnotesize
\begin{tabular}{l|cc|cc|cc|cc}
\toprule
  \textbf{Development set}                    & \multicolumn{2}{c|}{Libri-female}  & \multicolumn{2}{c|}{Libri-male}    & \multicolumn{2}{c|}{VCTK-diff-female} & \multicolumn{2}{c}{VCTK-diff-male} \\ \cline{1-9} 
Anony. Type             & \multicolumn{1}{c|}{OA}    & AA    & \multicolumn{1}{c|}{OA}    & AA    & \multicolumn{1}{c|}{OA}      & AA     & \multicolumn{1}{c|}{OA}     & AA    \\ \cline{1-9} 
VPC2020 B1 \cite{tomashenko2020introducing}             & \multicolumn{1}{c|}{50.14} & 36.79 & \multicolumn{1}{c|}{57.76} & 34.16 & \multicolumn{1}{c|}{49.97}   & 26.11  & \multicolumn{1}{c|}{53.95}  & 30.92 \\ 
S-ECAPA + HuBERT  & \multicolumn{1}{c|}{27.84} & 21.59  & \multicolumn{1}{c|}{7.76} & 13.66 & \multicolumn{1}{c|}{28.52}   & 11.68  & \multicolumn{1}{c|}{23.28}  & 16.18  \\ 
S-ECAPA + HuBERT-km& \multicolumn{1}{c|}{49.43} & 37.38 & \multicolumn{1}{c|}{48.45} & 39.29 & \multicolumn{1}{c|}{51.15}   & 26.78  & \multicolumn{1}{c|}{49.68}  & 30.47 \\
S-ECAPA + HuBERT-soft   & \multicolumn{1}{c|}{48.15} & 35.65  & \multicolumn{1}{c|}{43.32} & 35.71 & \multicolumn{1}{c|}{51.09}   & 27.40  & \multicolumn{1}{c|}{48.24}  & 30.52  \\ 
F-ECAPA + HuBERT-soft & \multicolumn{1}{c|}{47.44} & 29.55 & \multicolumn{1}{c|}{46.72} & 34.78 & \multicolumn{1}{c|}{52.11}   & 24.09  & \multicolumn{1}{c|}{48.04}  & 29.48 \\ 
\bottomrule
\end{tabular}
\vspace{-5mm}
\end{table*}

\begin{table*}[htbp]
  \centering
  \footnotesize
  \caption{
   EER (\%) on English data (test set of VPC 2020) evaluated by \emph{$ASV_\text{eval}^{\textrm{\textrm{eng}}}$}. 
  The EER results of the OO condition for Libri-female, Libri-male, VCTK-diff-female, and VCTK-diff-male are 7.66\%, 1.11\%, 4.88\%, and 2.06\%, respectively. Higher EER indicates better privacy.}
\vspace{1mm}
\label{tab:vpc-test-eer}
\setlength{\tabcolsep}{12pt}
 \begin{tabular}{l|cc|cc|cc|cc}
\toprule
 \textbf{Test set}                    & \multicolumn{2}{c|}{Libri-female}  & \multicolumn{2}{c|}{Libri-male}    & \multicolumn{2}{c|}{VCTK-diff-female} & \multicolumn{2}{c}{VCTK-diff-male} \\ \hline
Anony. Type             & \multicolumn{1}{c|}{OA}    & AA    & \multicolumn{1}{c|}{OA}    & AA    & \multicolumn{1}{c|}{OA}      & AA     & \multicolumn{1}{c|}{OA}     & AA    \\ \hline
VPC2020 B1 \cite{tomashenko2020introducing}              & \multicolumn{1}{c|}{47.26} & 32.12 & \multicolumn{1}{c|}{52.12} & 36.75 & \multicolumn{1}{c|}{48.05}   & 31.74  & \multicolumn{1}{c|}{53.85}  & 30.94 \\
S-ECAPA + HuBERT  & \multicolumn{1}{c|}{20.99} & 14.42  & \multicolumn{1}{c|}{12.92} & 14.52 & \multicolumn{1}{c|}{26.03}   & 16.72  & \multicolumn{1}{c|}{17.57}  & 16.30  \\ 
S-ECAPA + HuBERT-km& \multicolumn{1}{c|}{46.17} & 36.13 & \multicolumn{1}{c|}{44.99} & 38.98 & \multicolumn{1}{c|}{48.51}   & 33.64  & \multicolumn{1}{c|}{52.81}  & 32.61 \\
S-ECAPA + HuBERT-soft   & \multicolumn{1}{c|}{41.42} & 31.02  & \multicolumn{1}{c|}{39.87} & 36.97 & \multicolumn{1}{c|}{48.10}   & 31.22  & \multicolumn{1}{c|}{47.59}  & 35.94  \\ 
F-ECAPA + HuBERT-soft & \multicolumn{1}{c|}{41.24} & 29.74 & \multicolumn{1}{c|}{42.54} & 33.18 & \multicolumn{1}{c|}{50.31}   & 29.32  & \multicolumn{1}{c|}{48.11}  & 30.71 \\
\bottomrule
\end{tabular}
\vspace{-5mm}
\end{table*}

\subsection{Speaker Anonymization Experimental Setup}
\subsubsection{Evaluation Plan}
We followed the VPC 2020 evaluation plan \cite{tomashenko2020introducing,tomashenko2021voiceprivacy} for our speaker anonymization experiments. 
The only difference is that the content encoders of the proposed systems used SSL models pre-trained on an external database not included in the VPC 2020 protocol (i.e., \textit{LibriSpeech-train-960}). Details on the content encoders are explained in the next section..

The evaluation plan treated the speaker anonymization task as a game between users and attackers. It assumes that the attackers have access to an ASV model (which is referred to as \emph{$ASV_\text{eval}$}) and a few enrollment trials for each speaker. The attackers then use these resources to identify the speaker identity in anonymized test trials. Specifically, the evaluation plan defines three attack scenarios, and the enrollment and test trials in each scenario are either original (O) or anonymized (A). The three scenarios are illustrated in Figure \ref{fig:eval_types} and listed as follows:
\begin{itemize}
\item \textit{Unprotected (OO)}: no anonymization is applied, and attackers verify the original test trials against the original enrollment trials; 
\item \textit{Ignorant attacker (OA)}: users anonymize their trial data while attackers are unaware and simply use original enrollment data;
\item \textit{Lazy-informed (AA)}: users anonymize their trial data, and attackers partially know which anonymizer was used. Attackers use the same anonymizer to anonymize the enrollment data, hoping that it can be better linked with the anonymized trial data from the same target speaker. However, attackers do not know the detailed parameters and  anonymized the enrollment data using different pseudo-speakers. 
\end{itemize} 
The $ASV_\text{eval}$ EER in \textit{Unprotected (OO)} serves as the reference where no anonymization is applied. The EERs from the other two scenarios measure how well an anonymization system protects the speaker information in the test trials when facing attackers with varied amounts of prior knowledge on the anonymization system. 
Ideally, the EERs should be as high as 50\% in both OA and AA scenarios. 

The evaluation plan also uses the word error rate (WER) as a utility metric to measure how well the speech content is preserved after anonymization. The WER was computed using an ASR model (\emph{$ASR_\text{eval}$}). An ideal anonymization system should not increase the WER of test trials after anonymization. 
Note that both \emph{$ASR_\text{eval}$} and \emph{$ASV_\text{eval}$} are independent from the anonymization system. 
For the experiments on different languages, we use different \emph{$ASR_\text{eval}$} and \emph{$ASV_\text{eval}$} to evaluate the anonymization systems more accurately. Details are explained in the following sections.

Another utility metric is the naturalness of anonymized audios. We used a recently proposed mean opinion score (MOS) prediction network \cite{cooper2021generalization} to estimate perceived naturalness rather than conducting time-consuming listening tests. 

\subsubsection{System Configurations}
We built two versions of the proposed anonymization system. While both used the HuBERT-soft content encoder, they use either F-ECAPA or S-ECAPA as the speaker encoder. For reference, we built another two anonymization systems using S-ECAPA and slightly different HuBERT-based content encoders\footnote{We also tested wav2vec 2.0 as the content encoder and observed similar results to those using HuBERT. We thus only present the results of HuBERT-based systems in this paper.}.

The first reference system used a pre-trained HuBERT Base model from the Fairseq toolkit. 
Given an input waveform, this HuBERT model extracts a sequence of 768-dimensional vectors from the output of the sixth Transformer layer as the content representations \cite{polyak2021speech}. The second reference system used the HuBERT-km Base model released in the same toolkit. This model was based on HuBERT, but additional $k$-means clustering was applied to learn 200 clusters on \textit{LibriSpeech-train-clean-100}. During inference, each continuous-valued content representation was quantized, and the corresponding 200-dimensional cluster centroid was used as the content representation.
 
Finally, the HuBERT-soft in the proposed systems used HuBERT as the backbone and the index of the quantized content representation from HuBERT-km as the target. It was fine-tuned on \textit{LibriSpeech-train-clean-100} using the strategy in Section~\ref{sec:hubert soft}. Each content vector from HuBERT-soft has 200 dimensions.

All the anonymization systems used the same YAAPT F0 extractor.  Given the F0, content, and speaker vectors from the corresponding encoders, a HiFi-GAN was trained on \textit{LibriTTS-train-clean-100} for each anonymization system.
Noted we use cosine distance to generate the pseudo speaker unlike the B1 system using PLDA scores. 
\begin{table}[tbp]
\vspace{-3mm}
  \caption{ WER (\%) on English data evaluated by \emph{$ASR_\text{eval}^{\textrm{eng}}$}. Lower WER indicates better utility.}
  \vspace{1mm}
  \label{tab:asr-results}
  \centering
  \footnotesize
  \begin{tabular}{l|r|r|r|r}
\toprule
 & \multicolumn{2}{c|}{ Libri.} & \multicolumn{2}{c}{ VCTK} \\ \hline 
{Anonymization system} & {Dev.} & {Test} & {Dev.} & {Test} \\ \hline 
Ground Truth \cite{tomashenko2020introducing} & 3.83 & 4.15 & 10.79 & 12.82\\ 
VPC 2020 B1\cite{tomashenko2020introducing} & 6.39 & 6.73 & 15.38 & 15.23\\ 
S-ECAPA + HuBERT  & 4.23 & 4.47 & 12.12 & 13.89  \\
S-ECAPA + HuBERT-km & 7.84 & 7.80 & 19.21 &  21.74 \\
S-ECAPA + HuBERT-soft & 4.47 & 4.70 & 12.88 & 14.57\\
F-ECAPA + HuBERT-soft & 4.50 & 4.69 & 12.96 & 14.86\\
 \bottomrule
\end{tabular}
\vspace{-6mm}
\end{table}

\subsection{Speaker Anonymization Experiments in English}
\label{sec:English SA}
We evaluated the anonymization systems on the official development and test sets of VPC 2020. These two sets contain English utterances of several female and male speakers from the \textit{LibriSpeech} and VCTK \cite{yamagishi2019cstr} corpora. To compute the WER and EER,
we used the language-matched $ASR_\text{eval}$ and $ASV_\text{eval}$ systems provided by VPC 2020 \cite{tomashenko2020introducing,tomashenko2021voiceprivacy} and denoted them as $ASR_\text{eval}^{\textrm{eng}}$ and $ASV_\text{eval}^{\textrm{eng}}$, respectively. They were trained on the \textit{LibriSpeech-train-clean-360} English dataset. 

Tables \ref{tab:vpc-dev-eer} and \ref{tab:vpc-test-eer} show the EER results of various speaker anonymization systems, measured by \emph{$ASV_\text{eval}^{\textrm{eng}}$} on the VPC 2020 development and test sets, respectively. Table \ref{tab:asr-results} shows their WER results measured by \emph{$ASR_\text{eval}^{\textrm{eng}}$}.
VPC 2020 baseline B2 was not included in the objective evaluation as its objective results are worse than  B1\cite{tomashenko2021voiceprivacy}.
Here is the point-by-point summary: 

\noindent
\textbf{Proposed system vs.\ B1:} From the results, the first observation is that compared with VPC 2020 B1, our proposed systems, ``F-ECAPA + HuBERT-soft,'' and ``S-ECAPA + HuBERT-soft'' (the last two rows), which do not require any language-specific modules, have achieved comparable EER values on both OA and AA conditions and have lower WERs than those of B1.

\noindent
\textbf{Soft content encoder:} The WER and EER results of ``S-ECAPA + HuBERT'' show that although the continuous  representations extracted from ``S-ECAPA + HuBERT'' achieved low WERs, ASV EERs are far from 50\%. 
As expected, the continuous representations without $k$-means clustering do not have proper disentanglement of content and speaker, and this prevents the generation of properly speaker-anonymized speech.
Next, for ``S-ECAPA + HuBERT-km,'' we can see that the $k$-means clustering process managed to suppress speaker information and increased the ASV EERs of speaker-anonymized speech significantly whereas it also increased the ASR WERs, too. As hypothesized, the hard clustering process seems to cause pronunciation issues. 
Finally, we can see that the proposed ``S-ECAPA + HuBERT-soft'' system strikes a good balance and speaker-anonymized speech has a low ASR WER and high ASV EER.
\noindent
\textbf{S-ECAPA vs.\ F-ECAPA:}
Although not the main focus, it is also meaningful to compare acoustic features used for our speaker encoder.  Interestingly, ``S-ECAPA + HuBERT-soft'' always has higher EER values than ``F-ECAPA + HuBERT-soft'' on all AA conditions (where attackers have more priors). 
\begin{figure}[tb]
\vspace{-5mm}
\centering\includegraphics[width=70mm,height=50mm]{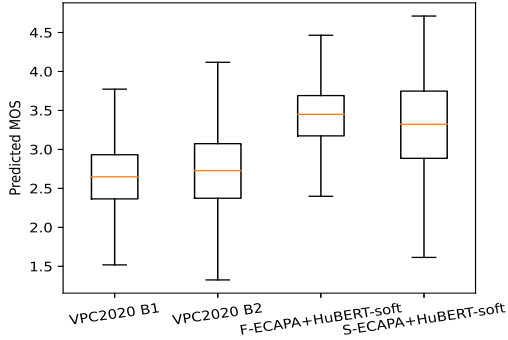}
\vspace{-3mm}
\caption{Box-plots on predicted naturalness scores of anonymized speech from experimental systems.}
\label{fig:vpc-mos}
\vspace{-5mm}
\end{figure}

To further analyze the effectiveness of our proposed models, we evaluate the naturalness of speaker-anonymized speech based on the MOS prediction network. Box-plots of the predicted MOS scores are shown in Figure \ref{fig:vpc-mos}. 
From the figure, we can observe a similar trend to the subjective results of VPC 2020 in which B2 had higher naturalness scores than B1 \cite{tomashenko2021voiceprivacy}. Moreover we can see that the proposed ``S-ECAPA + HuBERT-soft'' and ``F-ECAPA + HuBERT-soft'' are expected to have higher naturalness than the B1 and B2 systems. 

Overall, the proposed simpler ``ECAPA + HuBERT-soft'' anonymization systems without any language-specific models can protect the speaker information almost as well as the VPC 2020 primary baseline, and provide reliable linguistic information among all the anonymization systems.

\subsection{ Speaker Anonymization Experiments in Mandarin}
\label{sec:mand SA}
We directly used the proposed same anonymization systems from Section \ref{sec:English SA} on a Mandarin speaker anonymization task without training or fine tuning them on Mandarin data.
We evaluated the systems in the \textit{Unprotected (OO)}, \textit{Ignorant attacker (OA)}, \textit{Lazy-informed (AA)}, and a new scenario called \textit{OR}. In this {OR} scenario, the enrollment is original and the test trial is resynthesized by the anonymization system using the original speaker vector \cite{champion:hal-02995855}. An ideal system thus should not change speaker information or content of the test trial in the OR scenario, and the WER and EER should be as low as those from the {OO} scenario. 

The evaluation was conducted on a test set sampled from a multi-speaker Mandarin speech corpus called {AISHELL-3}\cite{shi21c_interspeech}. The test set contains 4,267 utterances from 44 speakers. We split them into test trial (88 utterances) and enrollment (4179 utterances) subsets, which are used to produce 10,120 enrollment-test trials for the ASV evaluation.
The ASV evaluation model \emph{$ASV_\text{eval}^{\textrm{mand}}$} was a F-ECAPA trained on Mandarin datasets called \textit{CN-Celeb-1 \& 2} \cite{li2022cn,fan2020cn}. 
\emph{$ASV_\text{eval}^{\textrm{mand}}$} was configured in the same way as Section \ref{sec:asv} described.
The ASR evaluation model \emph{$ASR_\text{eval}^{\textrm{mand}}$} 
was a publicly available ASR Transformer  \cite{speechbrain} trained on the 150-hour Mandarin ASR dataset {AISHELL-1} \cite{aishell_2017}. Note that the WER was replaced with character error rate (CER) in Mandarin ASR.

\begin{table}[tb]
\centering
\vspace{-5mm}
\caption{EER (\%) on Mandarin data evaluated by \emph{$ASV_\text{eval}^{\textrm{mand}}$}. Higher EER indicates better privacy.}
\vspace{1mm}
    \footnotesize
  \label{tab:asv-results-aishell-veri}
\begin{tabular}{l|c|c|c|c}
\toprule
EER(\%)                        & OO & OR & OA & AA\\ \hline           
F-ECAPA-HuBERT-soft & 2.04 & 9.13 & 37.58 & 22.98\\ 
S-ECAPA-HuBERT-soft & 2.04 & 13.45 & 40.81 & 23.26\\
\bottomrule
\end{tabular}
\vspace{-4mm}
\end{table}

\begin{table}[tb]
 \caption{CER (\%) on Mandarin data evaluated by \emph{$ASR_\text{eval}^{\textrm{mand}}$}. Lower CER indicates better utility.} 
 \vspace{1mm}
    \footnotesize
  \label{tab:asr-results-aishell-veri}
  \centering
\begin{tabular}{l|c|c }
\toprule
CER(\%)                & Syn. type           & Test set \\ \hline
{Ground Truth}        & -        & 10.36\\ 
{F-ECAPA + HuBERT-soft} & resyn. & 14.81 \\ 
{F-ECAPA + HuBERT-soft} & anony. &  18.86 \\
{S-ECAPA + HuBERT-soft} & resyn. &  14.95 \\
{S-ECAPA + HuBERT-soft}  & anony.& 21.18 \\ 
\bottomrule
\end{tabular}
\vspace{-4mm}
\end{table}

Tables \ref{tab:asv-results-aishell-veri} and \ref{tab:asr-results-aishell-veri} list the EERs and CERs, respectively. We first observe that both proposed systems increased the ASV EER to around 40\% in the OA scenario and 23\% in the AA scenario, which suggests that speaker information is protected to an acceptable degree. Meanwhile, the CERs on the anonymized trials were increased to about 20\%, suggesting degradation on the speech content. 
Interestingly, the CERs on anonymized trials are higher than those on resynthesized trials for both proposed systems.
The increased CER is likely to be caused by the pseudo x-vectors obtained from the English x-vector pool, while used to generate Mandarin speech. The previous study \cite{raj2019probing, williams2019disentangling} has proven that in addition to speaker-related information, x-vectors also encode speaker rate, background conditions and lexical content information. Fixing the data mismatch is one potential way to improve our proposed anonymization systems. 
Finally, experiments on Mandarin an English share a similar trend that compared with ``F-ECAPA + HuBERT-soft," ``S-ECAPA + HuBERT-soft" tends to sacrifice some linguistic content to protect the speaker information, obtaining a higher EER on OR, OA, and AA conditions but a worse CER on both resynthesis and anonymized trials. 

\section{Conclusion}
\label{sec:conclusion}
This paper proposed an SSL-based language-independent speaker anonymization method,
which consists of a HuBERT-based soft content encoder, ECAPA-TDNN speaker encoder, F0 encoder and HiFi-GAN decoder. 
Experiments on English VPC 2020 datasets and Mandarin AISHELL-3 datasets demonstrated that our proposed speaker anonymization method without any explicit language-specific resources,
can be adopted to other languages successfully. 
Future work will focus on building a more robust speaker anonymization method so that it can better process input trials recorded in an acoustic condition with unseen conditions.

\noindent 
\textbf{Acknowledgment}
This study is supported by JST CREST Grants (JPMJCR18A6 and JPMJCR20D3), MEXT KAKENHI Grants (21K17775, 21H04906, 21K11951, 18H04112), and the VoicePersonal project (ANR-18-JSTS-0001).

\bibliographystyle{IEEEbib}
\bibliography{ref}

\end{document}